\documentclass[twocolumn,aps,pra,showpacs,amsmath,superscriptaddress]{revtex4-1}
\usepackage{amsfonts}
\usepackage{amssymb}
\usepackage{mathrsfs}
\usepackage{graphicx}
\usepackage{float}
\usepackage{xcolor}
\usepackage{cases}
\usepackage[colorlinks,linkcolor=magenta,citecolor=blue,urlcolor=blue]{hyperref}
\usepackage{graphicx}
\usepackage{dcolumn}
\usepackage{bm}
\usepackage{amsmath}
\usepackage{amsbsy}
\usepackage{color}
\usepackage{epstopdf}
\usepackage{multirow}
\usepackage{etoolbox}

\makeatletter
\newcommand{\Rmnum}[1]{\expandafter\@slowromancap\romannumeral #1@}

\begin{document}
	\title{Exact multiple anomalous mobility edges in a flat band geometry}

	\author{Zhanpeng Lu}
	\thanks{These authors contributed equally to this work.}
	\affiliation{Institute of Theoretical Physics and State Key Laboratory of Quantum Optics Technologies and Devices, Shanxi University, Taiyuan 030006, China}
	\author{Hui Liu}
	\thanks{These authors contributed equally to this work.}
	\affiliation{Institute of Theoretical Physics and State Key Laboratory of Quantum Optics Technologies and Devices, Shanxi University, Taiyuan 030006, China}
    \author{Yunbo Zhang}
    \email{ybzhang@zstu.edu.cn}
    \affiliation{Zhejiang Key Laboratory of Quantum State Control and Optical Field Manipulation, Department of Physics, Zhejiang Sci-Tech University, Hangzhou 310018 , China}
    \author{Zhihao Xu}
	\email{xuzhihao@sxu.edu.cn}
	\affiliation{Institute of Theoretical Physics and State Key Laboratory of Quantum Optics Technologies and Devices, Shanxi University, Taiyuan 030006, China}
	\affiliation{Collaborative Innovation Center of Extreme Optics, Shanxi University, Taiyuan 030006, China}

	\begin{abstract}
		Anomalous mobility edges(AMEs), separating localized from multifractal critical states, represent a novel form of localization transition in quasiperiodic systems. However, quasi-periodic models exhibiting exact AMEs remain relatively rare, limiting the understanding of these transitions. In this work, we leverage the geometric structure of flat band models to construct exact AMEs. Specifically, we introduce an anti-symmetric diagonal quasi-periodic mosaic modulation, which consists of both quasi-periodic and constant potentials, into a cross-stitch flat band lattice. When the constant potential is zero, the system resides entirely in a localized phase, with its dispersion relation precisely determined. For non-zero constant potentials, we use a simple method to derive analytical solutions for a class of AMEs, providing exact results for both the AMEs and the system's localization and critical properties. Additionally, we propose a classical electrical circuit design to experimentally realize the system. This study offers valuable insights into the existence and characteristics of AMEs in quasi-periodic systems.
\end{abstract}

	\maketitle

\section{Introduction}
   {{Multifractal critical states} are special quantum states in disordered systems, characterized by strong amplitude fluctuations at the Anderson transition and displaying multifractal structures \cite{Huckestein,Jagannathan,Ostlund,Merlin,E,Ashraff,Evers}. Moreover, recently, {multifractal critical states} have been observed in cold atom systems using the momentum lattice technique \cite{Xiao2021SC}. Since the behavior of {multifractal critical states} is completely different from {either} extended or localized states \cite{XXCPL2024,Goblot2020,ZZhai2021}, they have attracted increasing attention from researchers over the past few decades. In terms of spatial distribution, the wave function of extended states is delocalized in real space but localized in the dual momentum space, whereas the opposite holds true for localized states \cite{Daiprb2024}.} This contrast suggests that extended and localized states can transform into each other under a position-momentum dual transformation. {Multifractal critical states}, in contrast, are delocalized in both real and momentum spaces \cite{LTarxiv2024,Lagendijk,LinPRB2023,SRoy2021PRL,Padhan2022PRB,ZWZPRA2022,SA2023PRB,SRoy2022PRB,TLT2021,Duncan2024,WYC2024,WYC2020}. {In position space, for an extended state, the amplitude between neighboring lattice sites remains constant in the thermodynamic limit, leading to a Lyapunov exponent $\lambda^{-1}=0$. In contrast, for a localized state, the amplitude decays exponentially between neighboring sites, resulting in $\lambda^{-1}>0$. In momentum space, extended states are characterized by $\lambda^{-1}>0$, while localized states correspond to $\lambda^{-1}=0$. However, multifractal critical states differ from both extended and localized states in that they exhibit zero Lyapunov exponents $\lambda^{-1}=0$ in both real space and momentum space, as emphasized in Refs \cite{XXCPL2024}.}  {With interactions, single-particle {multifractal critical states} can evolve into some novel many-body quantum phases, such as a many-body critical phase \cite{WYCLXJ2021} or a multifractal charge-density-wave order \cite{Ribeiro2}.} {Recently, an important progress of {multifractal critical states} is anomalous mobility
edges(AMEs) \cite{Yicaizhang2022,TL2023,XJL2023,LSZ2025,CYD2015}.} {Unlike traditional MEs, which separate extended and localized states \cite{Kraus2012PRL,Segev2013NP,Biddle2010PRL,Ganeshan2015PRL,XJL2022,YW2023,LZP2022,ZHXU2020,LLH2024,Zhihaoxu2021,Longhi2024,LT2020,LT2018,Longhi2024PRB,LuschenPRL,SarmaPRB2020,Bodyfelt2014,YaoPRL2020,HY2020PRL,GaoJun2,LZP2025FOP,LIUTONGCPB1,LIUTONGCPB2,EHCAPS1,EHCAPS2,paper1,paper2},} {AMEs} delineate energy boundaries that distinguish localized states from {multifractal critical states}, thereby broadening the scope of localization physics.

   On the other hand, {the study of flat band lattices has attracted widespread attention in multidisciplinary fields. Flat band lattices are translationally invariant systems that feature at least one dispersionless band and can be detected in phononic metamaterials \cite{Bilal2024PRL}. Due to the freezing of electron motion in flat band systems, they offer an ideal platform for achieving exotic correlated phases or peculiar transport phenomena \cite{Maksymenko2012,Rhim2021,Hyr2013,Derzhko2010,Goda2006,Huber2010,Green2010,FOPFLAT1,FOPFLAT2,FOPFLAT3}, such as boundary flat bands with topological spin textures \cite{Biao2024PRB}.} Recent theoretical studies have shown that introducing quasi-periodic modulations into flat band lattices can lead to exotic localization transitions, such as exact MEs \cite{Danieli2015}. However, although some research has investigated {AMEs} in flat band systems, there has been limited focus on the precise engineering of exact {AMEs} within these lattices. This gap presents a promising direction for future research in this largely unexplored area.

	{In this work, we investigate a one-dimensional flat-band cross-stitch lattice subjected to diagonal mosaic quasi-periodic modulation, which incorporates both a constant potential and a quasi-periodic potential. By leveraging the unique geometric structure of flat-band models, we construct exact AMEs in such systems. Specifically, we introduce an anti-symmetric diagonal quasi-periodic mosaic modulation into the cross-stitch lattice. Through a local rotational transformation, the system acquires a structure that supports AMEs, characterized by mosaic-type quasi-periodic hopping coefficients and on-site potentials. This framework enables the analytical derivation of exact AMEs in flat-band quasicrystalline systems. Distinct from conventional quasi-periodic models where the system typically transitions from an extended phase to an intermediate phase with MEs, our model exhibits fundamentally different behavior. In the absence of constant potential modulation, all eigenstates are localized and no AMEs are observed. However, when the constant potential modulation is present, the system exhibits a class of exact AMEs. For weak quasi-periodic modulation, the system remains in a critical phase. As the strength of the quasi-periodic modulation increases, the system transitions to an intermediate phase featuring multiple exact AMEs, with multifractal critical states persisting even in the strong disorder limit. Analytically, we demonstrate that the localization properties of the system can be mapped onto a generalized AA model with both diagonal and off-diagonal modulations, which exhibits the energy-dependent transitions from multifractal to localized states. This energy dependence underlies the emergence and persistence of AMEs in our study.} These findings reveal rich localization phenomena and provide new insights into the interplay between quasi-periodic modulation and AMEs in flat-band systems.
	
	The rest of the paper is organized as follows: In Sec.II, we first introduce the general model Hamiltonian of the 1D cross-stitch lattice with diagonal mosaic quasi-periodic modulation. In Sec.III, we study the localization properties of the system for the mosaic quasi-periodic modulated cross-stitch lattice. An experimental scheme using electrical circuits is proposed to realize our model in Sec. IV. Finally, a summary is presented in Sec. V.
	
	\section{Model}
	\begin{figure}[h]
	\includegraphics[width=0.5\textwidth]{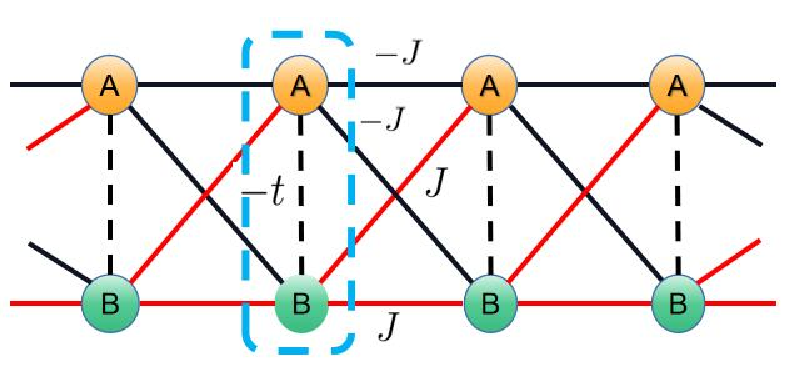}
	\caption{{Schematic diagram of the cross-stitch lattice. The blue dotted box region indicates the unit cell with two sublattices labeled $A$ and $B$.}}
	\label{Fig0}
\end{figure}
	We consider a cross-stitch lattice {\cite{Lee2023,Maimaiti2017,Maimaiti2021}} with an on-site mosaic quasi-periodic modulation. {The cross-stitch model, depicted in Fig.\ref{Fig0}, is defined for a unit cell comprising two sublattices, labeled $A$ and $B$, within the blue dotted box region.} Under the tight-binding approximation, the system is described by the Hamiltonian $\hat{H} = \hat{H}_0+\hat{H}^{\prime}$, where
	\begin{eqnarray}
		\hat{H}_0&=&-J\sum_{n}\left(\hat{c}^{\dag}_{n,A}\hat{c}_{n+1,A}+\hat{c}^{\dag}_{n,A}\hat{c}_{n+1,B}+H.c\right)\notag\\
		&&+J\sum_{n}\left(\hat{c}^{\dag}_{n,B}\hat{c}_{n+1,B}+\hat{c}^{\dag}_{n,B}\hat{c}_{n+1,A}+H.c\right) \notag \\
		&&-t\sum_{n}\left(\hat{c}^{\dag}_{n,A}\hat{c}_{n,B}+H.c\right),
		\label{eq1}
	\end{eqnarray}
	and
	\begin{eqnarray}
		\hat{H}^{\prime}=\sum\limits_{n}\left(V_{A,n}\hat{c}^{\dag}_{n,A}\hat{c}_{n,A}+V_{B,n}\hat{c}^{\dag}_{n,B}\hat{c}_{n,B}\right),
		\label{eq2}
	\end{eqnarray}
	where $\hat{c}_{n,A}$($\hat{c}_{n,B}$) is the annihilation operator localized at sublattice $A$($B$) of the $n$-th cell, $J$ and $t$ represent the intercell and intracell hopping amplitudes, respectively. Here, we set $J=1$ as the unit of energy. {The on-site potentials on atoms $A$ and $B$ are considered to have an anti-symmetric form, i.e., $V_{A,n}=-V_{B,n}=V_{n}$.} The on-site mosaic modulation {$V_{n}$} is defined as {\cite{XJL2023,Longhi2024PRB,HY2020PRL}}
	\begin{equation}
		V_{n}= \begin{cases}
			\Delta_1\cos(2\pi\beta n) &  n=\kappa s,\\
			\Delta_2 & \text{otherwise},  \\
		\end{cases}\label{eq3}
	\end{equation}
	where the spatial modulation frequency set as the inverse of the golden mean, $\beta=(\sqrt{5}-1)/2$, $\Delta_1$ and $\Delta_2$ denote the quasi-periodic and constant potential amplitudes, respectively, and $\kappa$ is an integer. Since the quasi-periodic potential appears periodically with an interval of $\kappa$ for the unit cell in the cross-stitch chain, we group the nearest $\kappa$ unit cells into a cluster, indexed by $s$. If the total number of the cluster is $\tilde{N}$, the system size will be $L=2\kappa\tilde{N}$ with the number of unit cells $N=\kappa\tilde{N}$. When $\kappa=1$, $\hat{H}^{\prime}$ reduces to the standard AA potential. This AA-type modulated cross-stitch chain exhibits a critical-to-insulator transition as the strength of the quasi-periodic potentials increases at $\Delta_1=4$ without showing any {AMEs} \cite{Lee2023}. For $\kappa\neq1$ cases, we find that the localization properties strongly depend on whether $\Delta_2$ is present or absent.
	
	The eigenvalue problem for the given tight-binding model is formulated as follows:
	\begin{eqnarray}
		E\psi_n&=&-\hat{t}\psi_n-\hat{T}\psi_{n-1}-\hat{T}^{\dag}\psi_{n+1}+\hat{V}\psi_n,
		\label{eq4}
	\end{eqnarray}
	where
	\begin{equation}
		\hat{t}=\begin{pmatrix} 0 & t \\ t & 0 \end{pmatrix}, \quad \hat{T}=\begin{pmatrix}	1 & -1 \\ 1 & -1 \end{pmatrix}, \quad \hat{V}=\begin{pmatrix} V_n & 0 \\ 0 & -V_n \end{pmatrix}.
		\label{eq5}
	\end{equation}
	Here, $\hat{t}$, $\hat{T}$, and $\hat{V}$ represent the intracell coupling, intercell coupling, and on-site potential matrices, respectively. The vector $\psi_n=\left(a_n\ b_n\right)^{\rm{T}}$ corresponds to the two-site wavefunction in the $n$-th unit cell of the periodic lattice. In the crystalline limit, where the on-site potential is set to zero, the system exhibits two flat bands with eigenvalues $E_{\pm}=\pm\sqrt{t^2+4}$ \cite{Maimaiti2017,Maimaiti2021}. For such flat-band system, the lattice can be transformed into a Fano defect configuration using a local coordinate transformation\cite{Danieli2015}. Specifically, for the cross-stitch lattice, the transformation involves the matrix $\hat{U}$:
	\begin{equation}
		\begin{pmatrix} p_n  \\ f_n \end{pmatrix}=\hat{U}\begin{pmatrix} a_n \\ b_n \end{pmatrix}, \quad
		\hat{U}=\frac{1}{\sqrt{2}}\begin{pmatrix} 1 & 1 \\ 1 & -1 \end{pmatrix}.
		\label{eq6}
	\end{equation}
	After this transformation, the eigenvalue equation becomes:
	\begin{eqnarray}
		(E+t)p_n&=&-2f_{n-1}+V_nf_n, \notag\\
		(E-t)f_n&=&-2p_{n+1}+V_np_n.
		\label{eq7}
	\end{eqnarray}
	Eliminating $f_n$ in favor of $p_n$, we arrive at a reduced tight-binding form for $p_n$:
	\begin{eqnarray}
		B_0p_n=-2V_{n-1}p_{n-1}-2V_{n}p_{n+1}+V^2_{n}p_{n}
		\label{eq8}
	\end{eqnarray}
	with $B_0=E^2-t^2-4$. This formulation described a 1D mosaic-modulated chain, referred to as the P-chain. Given the quasiperiodic modulation with an interval $\kappa$, we can define a quasicell that includes the nearest $\kappa$ sites of the P-chain. The total number of the quasicell is denoted by $\tilde{N}$, and the system size of the P-chain is given by $N=\kappa\tilde{N}$.
	
	To investigate the localization properties of the eigenstates, one can calculate the inverse participation ratio (IPR) \cite{Evers}, $\mathrm{IPR}^{(i)}=\sum_{n=1}^{N}\sum_{\alpha=\{A,B\}}|\psi_{n,\alpha}^{(i)}|^4$, where $\psi_{n,\alpha}^{(i)}$ represents the probability amplitude of the $i$-th eigenstate, with eigenvalue $E_i$, on the $\alpha$ sublattice in the $n$-th unit cell. For an extended state, the IPR scales as the inverse of the system size and approaches zero in the thermodynamic limit. Conversely, for a localized state, the IPR remains finite and independent of the system size $L$. In regions where the eigenstate $\psi^{(i)}$ exhibits multifractal behavior, $\mathrm{IPR}^{(i)}\propto L^{-\Gamma_i}$ with $\Gamma_i\in(0,1)$. To quantify this behavior, the fractal dimension of the $i$-th eigenstate $\psi_{n,\alpha}^{(i)}$ is defined as \cite{WYC2020}:
	\begin{equation}
		\mathrm{\Gamma}_i = -\lim_{L\to\infty}\left[\frac{\ln\mathrm{IPR}^{(i)}}{\ln{L}}\right].
		\label{eq9}
	\end{equation}
	Based on this definition, we can know that $\Gamma_i\to 1$ for an extended state; $\Gamma_i\to 0$ for a localized state; and when $\Gamma_i\in(0,1)$, the corresponding state is a {multifractal critical} one. Additional, the mean inverse participation ratio can be used to characterize the localization properties within a specific region,  $\mathrm{MIPR}(\tilde{\sigma}_{\tilde{\beta}})=(1/\mathcal{N}_{\tilde{\sigma}_{\tilde{\beta}}})\sum_{i\in{\tilde{\sigma}_{\tilde{\beta}}}}\mathrm{IPR}^{(i)}$, where $\tilde{\beta}\in\{\tilde{E},\tilde{C},\tilde{L}\}$ specifies the region type--extended ($\tilde{\sigma}_{\tilde{E}}$), critical ($\tilde{\sigma}_{\tilde{C}}$), or localized ($\tilde{\sigma}_{\tilde{L}}$)--and $\mathcal{N}_{\tilde{\sigma}_{\tilde{\beta}}}$ denotes the total number of eigenstates in the corresponding region. The scaling behavior of MIPR can be fitted as $\mathrm{MIPR}(\tilde{\sigma}_{\tilde{\beta}})\propto L^{-\tilde{\Gamma}}$. For a perfectly localized region, $\mathrm{MIPR}(\tilde{\sigma}_{\tilde{L}})$ maintains a finite value that hardly changes with the system size. For a perfectly extended region, $\mathrm{MIPR}(\tilde{\sigma}_{\tilde{E}})$ changes linearly with $1/L$. For a critical region, $\tilde{\Gamma} \in (0,1)$. This framework provides a method to analyze the localization and multifractal properties of eigenstates across various spectral regions.
	
	To further verify the presence of {multifractal critical states} in the system, we analyze the standard deviations of the coordinates ($\sigma_i$) and the inverse of localization length ($\lambda^{-1}_i$) of the eigenstates as functions of their eigenvalues. The standard
	deviations $\sigma_i$, as defined in \cite{Yicaizhang2022}, captures the spatial distribution of the eigenstate and is given by:
	\begin{equation}
		\sigma_i=\sqrt{\sum\limits_{j=1}^{L}\left(j-\overline{j}\right)^2\left|\psi^i(j)\right|^2,}
		\label{eq10}
	\end{equation}
	where $j$ represents the lattice coordinate, and $\overline{j}=\sum_{j=1}^{L}j\left|\psi^{(i)}(j)\right|^2$ is the center of mass position of the eigenstate. The value of $\sigma_i$ provides the information about the spatial distribution of the eigenstates. For localized (extended) states, $\sigma_i$ is relatively small (large). However, for {multifractal critical states}, $\sigma_i$ lies between these two extremes and typically exhibits fluctuations. For a given energy $E_i$, the inverse of localization length $\lambda^{-1}_i$, also referred to as the Lyapunov exponent \cite{Avila,You,Shu,Zhou}, is another critical metric for characterizing eigenstates. It is defined as \cite{WZH2022PRB,MacKinnon1983,yanxialiu2}:
	\begin{equation}
		\lambda_i^{-1} = \lim_{N\to\infty}\frac{1}{N}\ln||T(E_i)||,
		\label{eq11}
	\end{equation}
	where $\|T(E_i) \|$ denotes the norm of the total transfer matrix, $T(E_i)=\prod_{n=1}^{N}T_n(E_i)$. According to Eq. (\ref{eq8}), the transfer matrix for the $n$-th unit cell is expressed as:
	\begin{equation}
		T_n(E_i)=\begin{pmatrix}
			-\frac{\left(E_i^2-t^2-4-V_n^2\right)}{2V_n} & -\frac{V_{n-1}}{V_n} \\
			1 & 0\end{pmatrix}.
		\label{eq12}
	\end{equation}
	Here, $\lambda_i^{-1}$ measures the average growth rate of the wavefunction for the $i$-th eigenstate. A localized state corresponds to $\lambda_i^{-1}>0$, indicating exponential decay of the wavefunction. A delocalized state, whether extended or {multifractal critical}, is characterized by $\lambda_i^{-1}=0$. By examining the behaviors of $\sigma_i$ and $\lambda_i^{-1}$, one can distinguish between localized, extended and {multifractal critical states}.
		\begin{figure}[h]
		\includegraphics[width=0.5\textwidth]{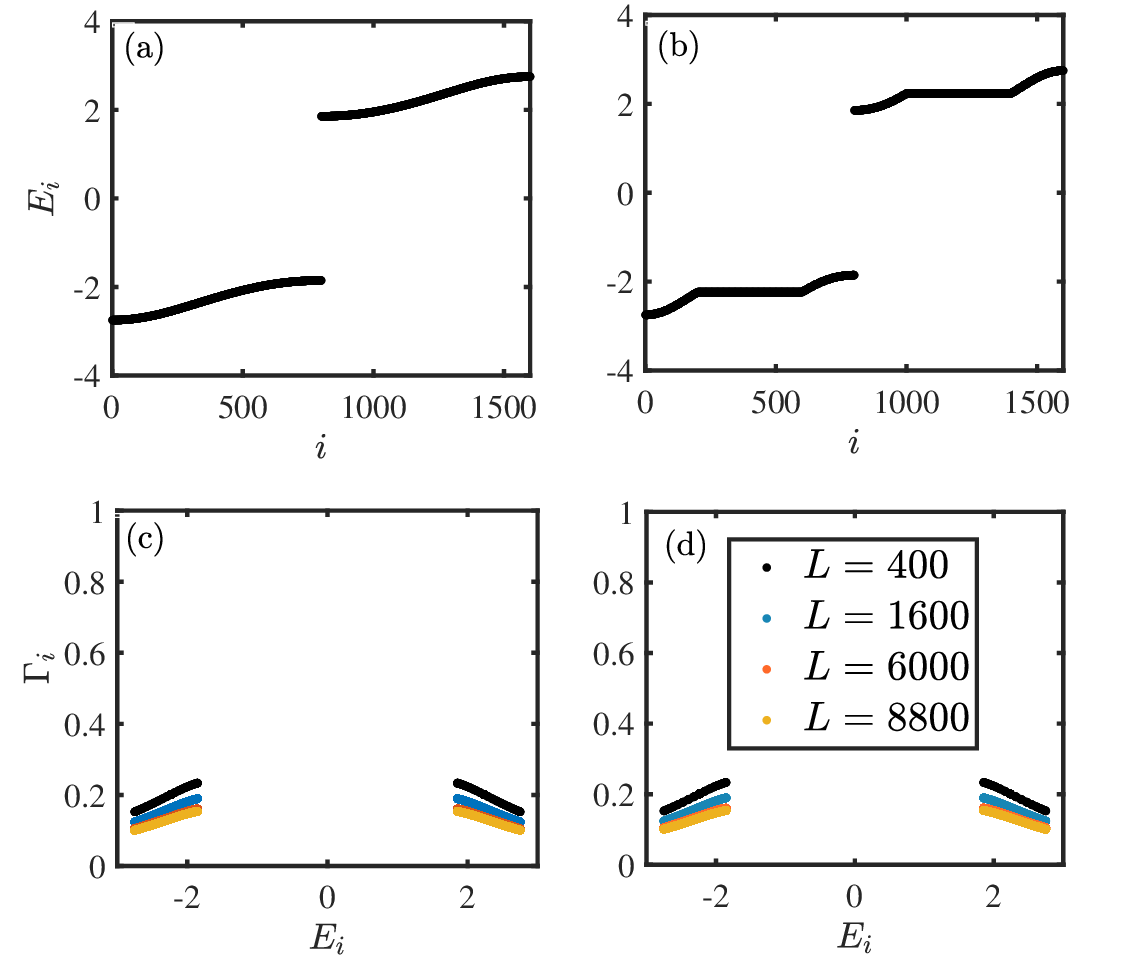}
		\caption{The energy distributions with $t=1$, $\Delta_1=1$, and $L=1600$ for (a) $\kappa=2$ and (b) $\kappa=4$, respectively. The fractal dimension $\Gamma_i$ versus $E_i$ with $t=1$, $\Delta_1=1$, and different system sizes for (c)$\kappa=2$ and (d)$\kappa=4$, respectively.}
		\label{Fig1}
	\end{figure}	
	In this paper, we apply the exact diagonalization method to numerically calculate the modulated cross-stitch chain under periodic boundary conditions (PBCs), with the intracell amplitude set to $t=1$.

	\section{Localization features}

	Let us first examine the case where $\Delta_2=0$. For an arbitrary $\kappa\ge 2$, the absence of $\Delta_2$ leads Eq. (\ref{eq8}) to describe {a series of disconnected lattice sites}, which implies that all eigenstates are localized. Moreover, there exists a $(\kappa-2)\tilde{N}$-fold degeneracy at $E=\pm\sqrt{t^2+4}$, respectively. The degeneracy of the remaining $L-2(\kappa-2)\tilde{N}$ energies is broken, and these energies are given by
	\begin{equation}
		E=\pm\sqrt{t^2+4+V_{\kappa s}^2/2\pm\sqrt{4V_{\kappa s}^2+V_{\kappa s}^4/4}}.
		\label{eq13}
	\end{equation}
	Figures \ref{Fig1}(a) and \ref{Fig1}(b) show the energy distributions with $t=1$, $\Delta_1=1$, and $L=1600$ for $\kappa=2$ and $\kappa=4$, respectively. In the case of $\kappa=2$, there are no degenerate states, while for $\kappa=4$, there is a $400$-fold degeneracy at $E=\pm\sqrt{5}$, respectively. The non-degenerate energies in Figs. \ref{Fig1}(a) and \ref{Fig1}(b) correspond to the values predicted by Eq. (\ref{eq13}), which agree well with our numerical results. Due to the emergence of the macroscopic degeneracy, arbitrary linear combinations of the degenerate eigenstates become solutions to the system. To mitigate the effects of this degeneracy, we introduce a small random disorder at each site in our numerical calculations. Specifically, we use a random disorder amplitude set to $10^{-8}$ in Figs. \ref{Fig1}(c) and \ref{Fig1}(d), which plot the fractal dimension $\Gamma_i$ as a function of eigenvalues for different system sizes for $\kappa=2$ and $\kappa=4$, respectively. Notably, $\Gamma_i$ approaches $0$ for all states as the system size increases, indicating that the system remains in a localized phase in the $\Delta_2=0$ case.
	
	For the case {of} $\Delta_2\neq 0$, the system exhibits distinct localization properties, and the {AMEs} emerge. We choose $\kappa=2$ and $3$ as examples for our discussion. Results for arbitrary $\kappa\ge 2$ are provided in the Appendix.
	
	\begin{figure}[h]
		\includegraphics[width=0.52\textwidth]{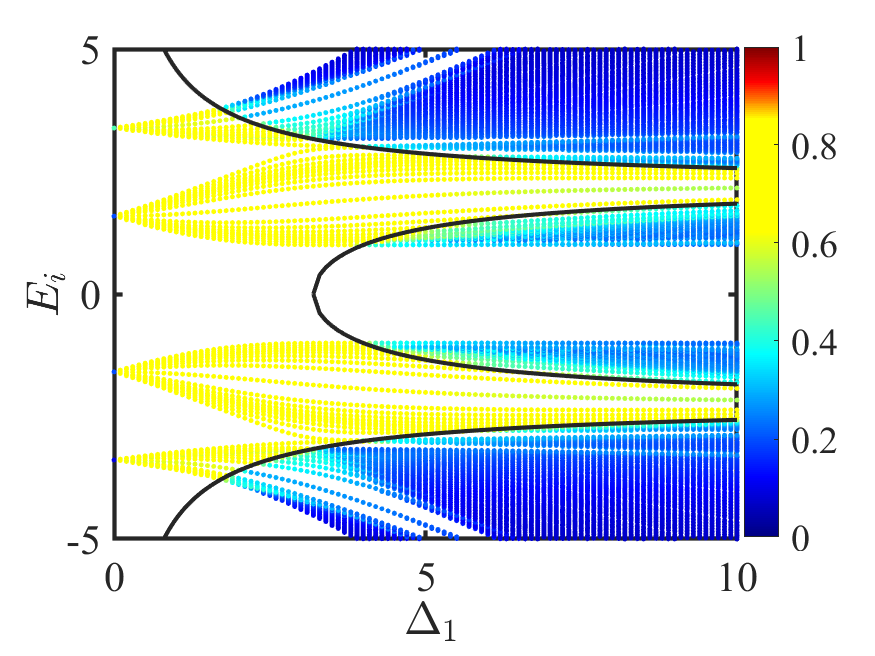}
		\caption{The fractal dimension $\Gamma_i$ of different eigenstates as a function of the corresponding $E_i$ and $\Delta_1$ for {$L=1220$} and $\Delta_2=2$. The {black} lines represent the {AMEs} given in Eq. (\ref{eq19}).}
		\label{Fig2}
	\end{figure}
	\begin{figure}[h]
		\includegraphics[width=0.5\textwidth]{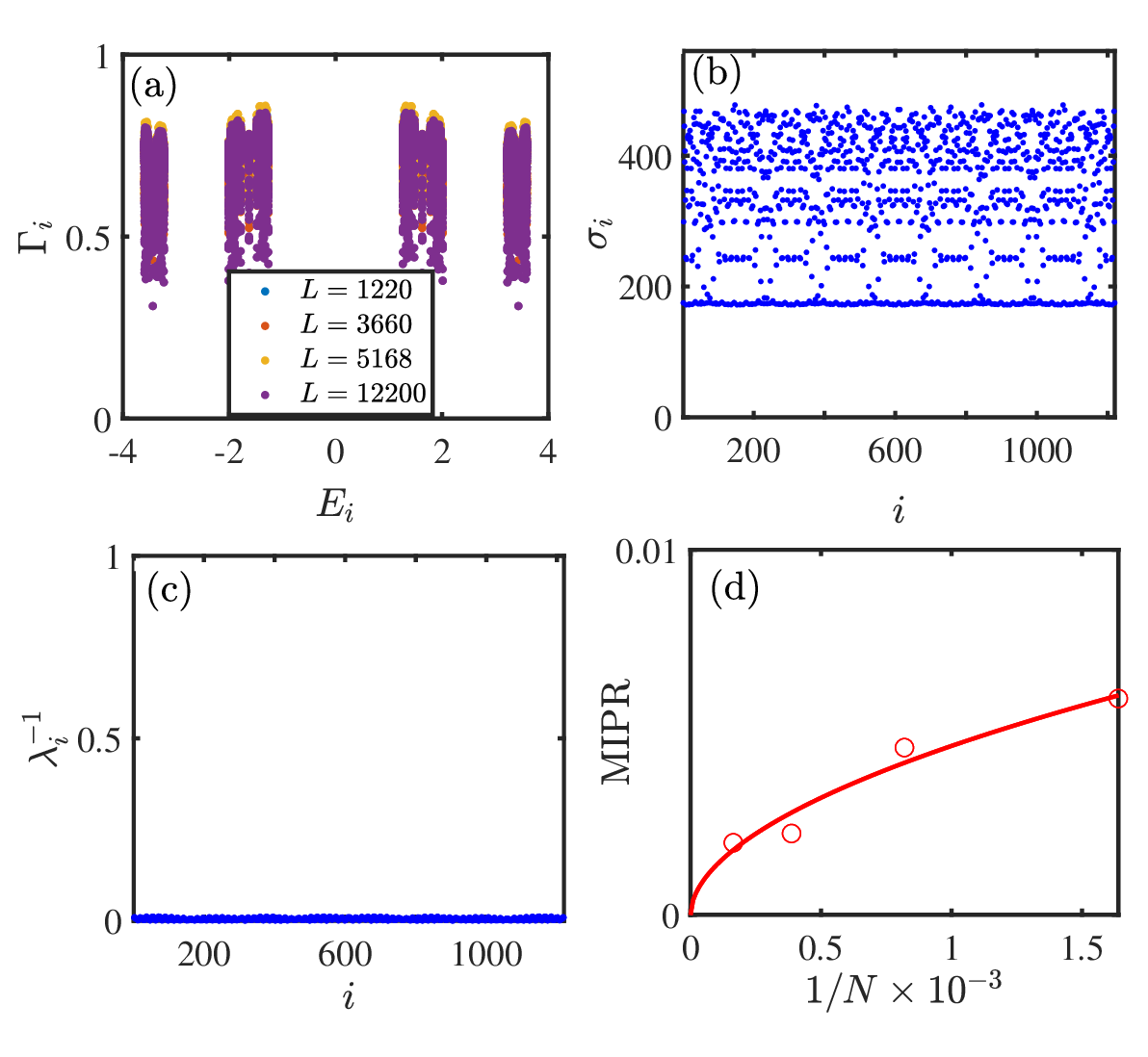}
		\caption{{{Multifractal critical states} in the weak disorder regime.} (a)The fractal dimension $\mathrm{\Gamma}_i$ versus $E_i$ for different system sizes. {(b,c)} The distributions of the eigenstate coordinates $\sigma_i$ and Lyapunov exponent $\lambda^{-1}_i$ as functions of the energy index. (d)The \rm{MIPR} as a function of $1/N$. Here, $\kappa=2$, $\Delta_1=1$, $\Delta_2=2$, and $L=1220$ for (b) and (c).}
		\label{Fig3}
	\end{figure}
	
	For $\kappa=2$, the quasiperiodic potential periodically appears at even unit cells, while the modulation amplitude at the odd unit cells is set to a constant value $\Delta_2 \ne 0$. From Eq. (\ref{eq8}), the effective eigenvalue equations for the P-chain can be expressed as:
	\begin{align}
		A_1p_{1,s}&=-2V_{2,s-1}p_{2,s-1}-2\Delta_2p_{2,s} \notag\\
		B_0p_{2,s}&=-2\Delta_2p_{1,s}-2V_{2,s}p_{1,s+1}+V_{2,s}^2p_{2,s} \label{eq17},
	\end{align}
	where $A_1=B_0-\Delta_2^2$. Here, the subscripts of $p_{\tilde{\alpha},s}$ and $V_{\tilde{\alpha},s}$ denote the $\tilde{\alpha}$-th site in the $s$-th quasicell, with $\tilde{\alpha}=1,2,\cdots,\kappa$ and $s=1,2,\cdots,\tilde{N}$. By reducing Eq.(\ref{eq17}), we obtain
	\begin{eqnarray}
		\left(B_0-\frac{4\Delta_2^2}{A_1}\right)p_{2,s}&=&\frac{4\Delta_2}{A_1}\left(V_{2,s-1}p_{2,s-1}+V_{2,s}p_{2,s+1}\right)\notag\\
		&&+\left(\frac{4}{A_1}+1\right)V_{2,s}^2p_{2,s}.
		\label{eq18}
	\end{eqnarray}
	This eigenvalue equation Eq. (\ref{eq18}) represents a 1D generalized AA model with quasiperiodically modulated hopping and on-site terms. The system undergoes a critical-to-insulator transition when the on-site amplitude reaches twice the hopping amplitude, as discussed in \cite{WYCLXJ2021,CYD2015,Lee2023}. Based on the localization transition of the generalized AA model, the analytical expressions for the {AMEs} are given by:
	\begin{equation}
		E_c=\pm\sqrt{\pm\frac{8\Delta_2}{\Delta_1}+t^2+\Delta_2^2},
		\label{eq19}
	\end{equation}
	which are shown as the {black} solid lines in Fig.\ref{Fig2}. Figure \ref{Fig2} illustrates the fractal dimension $\Gamma_i$ as the function of eigenenergy $E_i$ and $\Delta_1$ for {$L=1220$}. {The yellow and blue regions correspond to critical and localized states, respectively.} When the modulation amplitude $\Delta_1$ is weak, $\Gamma_i$ for all the eigenstates deviates from $0$ and $1$, indicating that the system resides in a critical phase. As $\Delta_1$ increases to $1.65$, a pair of {AMEs} emerges, marking the transition from the critical phase to a mixed phase of critical and localized states. When $\Delta_1\approx 4.1$, another pair of {AMEs} emerges. In the regime of large $\Delta_1$, there are four {AMEs} in total, which separate critical and localized states. {Notably}, {multifractal critical states} persist even in regions with large disorder, highlighting the robustness of these states.
	
	Figure \ref{Fig3}(a) shows the fractal dimensions $\Gamma_i$ as a function of eigenvalues for different system sizes in the weak disorder regime ($\Delta_1=1$). In this regime, the fractal dimensions for all states deviate from both $0$ and $1$ and exhibit minimal dependence on system size, indicating that all states are critical. Figures \ref{Fig3}(b) and \ref{Fig3}(c) show $\sigma_i$ and the Lyapunov exponent $\lambda_i^{-1}$ as functions of the energy index for $\Delta_1=1$, respectively. In Fig. \ref{Fig3}(b), the large fluctuations in $\sigma_i$ suggest the presence of {multifractal critical states}. Furthermore, as seen in Fig. \ref{Fig3}(c), $\lambda_i^{-1}$ approaches zero, confirming the delocalized nature of the eigenstates in the weak $\Delta_1$ regime. Additionally, the critical phase is explicitly confirmed by the scaling behavior of the MIPR. For $\Delta_1=1$, all states are critical, and the corresponding fitting function is ${\rm{MIPR}(\tilde{\sigma}_{\tilde{C}})} \approx 0.1789 N^{-0.5291}$, as shown in Fig. \ref{Fig3}(d).
	
	\begin{figure}[h]
		\includegraphics[width=0.53\textwidth]{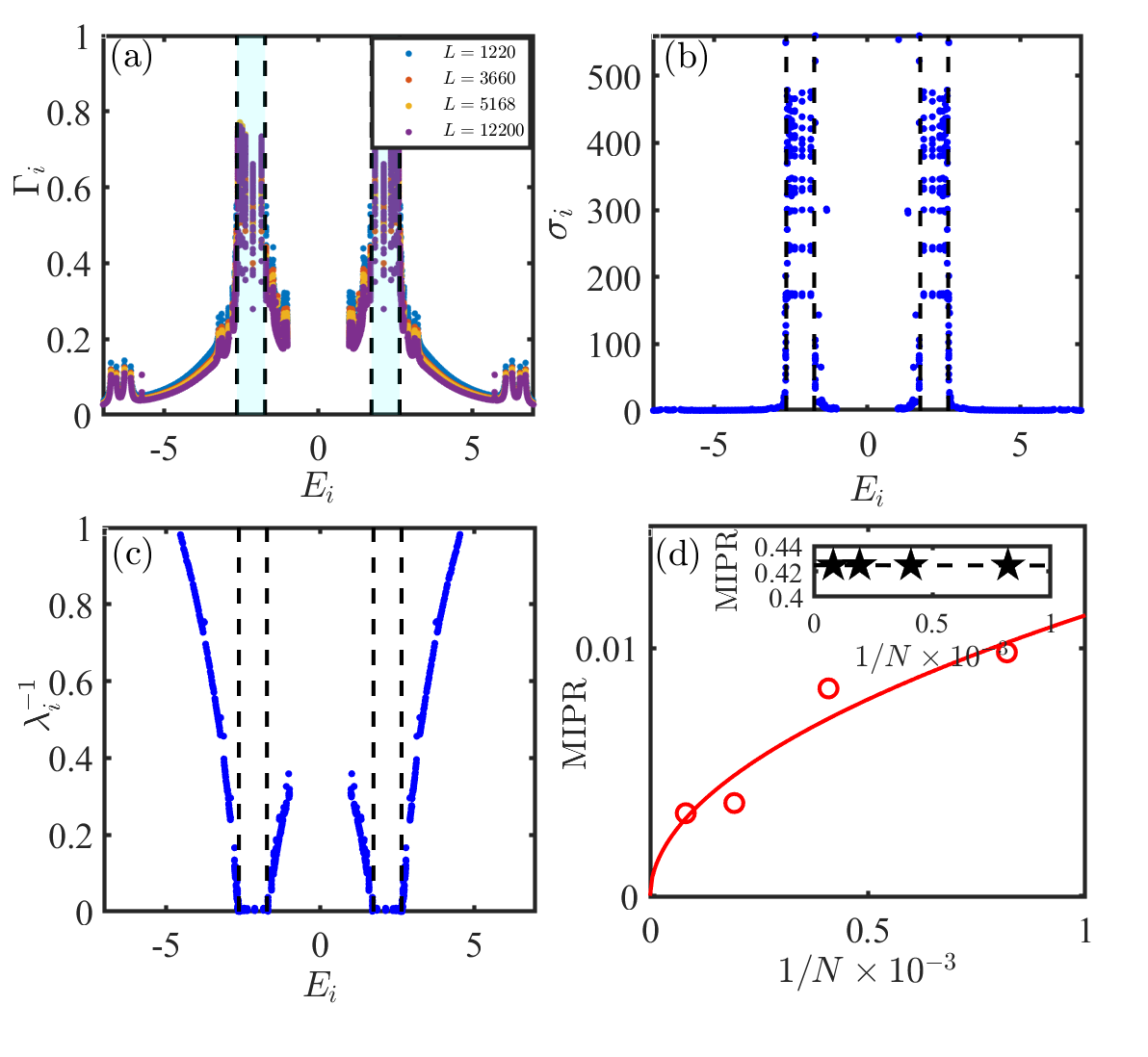}
		\caption{(a) The fractal dimension $\mathrm{\Gamma}_i$ versus $E_i$ for different system sizes. The shaded regions represent the critical regimes. (b) $\sigma_i$ as a function of {$E_i$}. (c) The Lyapunov exponent $\lambda^{-1}_i$ as a function of {$E_i$}. The black dash lines represent eigenstate indices of the {AMEs} given by Eq.(\ref{eq19}). (d) The MIPR for the critical regions as a function $1/N$. The insets show the scalings of the MIPR for the corresponding localized regions. Here, $\Delta_1=8$ and $\Delta_2=2$.}
		\label{Fig4}
	\end{figure}

	{Figure \ref{Fig2} suggests multiple {AMEs} emerge when $\Delta_1>1.65$.} Figures \ref{Fig4}(a) display the fractal dimension $\mathrm{\Gamma}_i$ of various eigenstates as a function of their corresponding eigenvalues for different system sizes with large modulation amplitude $\Delta_1=8$. The black dashed lines mark the positions of the {AMEs}, determined by Eq.(\ref{eq19}). In Fig. \ref{Fig4}(a), the four {AMEs} partition the energy band into for five regions. The shade areas represent regions with system-size-independent $\Gamma_i$, indicating the emergence of {multifractal critical states}. In constrast, the remaining regions are localized, with $\Gamma_i\to 0$ as the system size increases. The standard deviation $\sigma_i$ exhibits significant fluctuations in the critical regions, further confirming their critical nature, as shown in the Figs. \ref{Fig4}(b). Additionally, the corresponding values of $\lambda^{-1}_i$ approach $0$ in these critical regions, indicating the delocalized behavior, as seen in the Figs. \ref{Fig4}(c). Conversely, in the localized regions, $\lambda^{-1}_i>0$, and $\sigma_i$ is very small, reflecting localized properties. The coexistence of both {multifractal critical} and localized states is a hallmark {feature} of the intermediate phase {hosting} {AMEs}. Based on the exact expressions for {AMEs} in Eq. (\ref{eq19}) with $\kappa=2$, we can distinguish the critical and localized regions in the spectrum. We further calculate the MIPR for a specific region as a function of $1/N$, as shown in Fig \ref{Fig4}(d). One can see that the scaling of the MIPR for the two critical regions $\mathrm{MIPR}(\tilde{\sigma}_{\tilde{C}})\approx0.2779N^{-0.515}$ also exhibits critical behavior. The scalings of the MIPRs in the localized regions are shown in the insets of Fig. \ref{Fig4}(d). In these regions, the system-size-independent $\mathrm{MIPR}(\tilde{\sigma}_{\tilde{L}})$ tend to finite values in the thermodynamic limit.

	\begin{figure}[h]
		\includegraphics[width=0.48\textwidth]{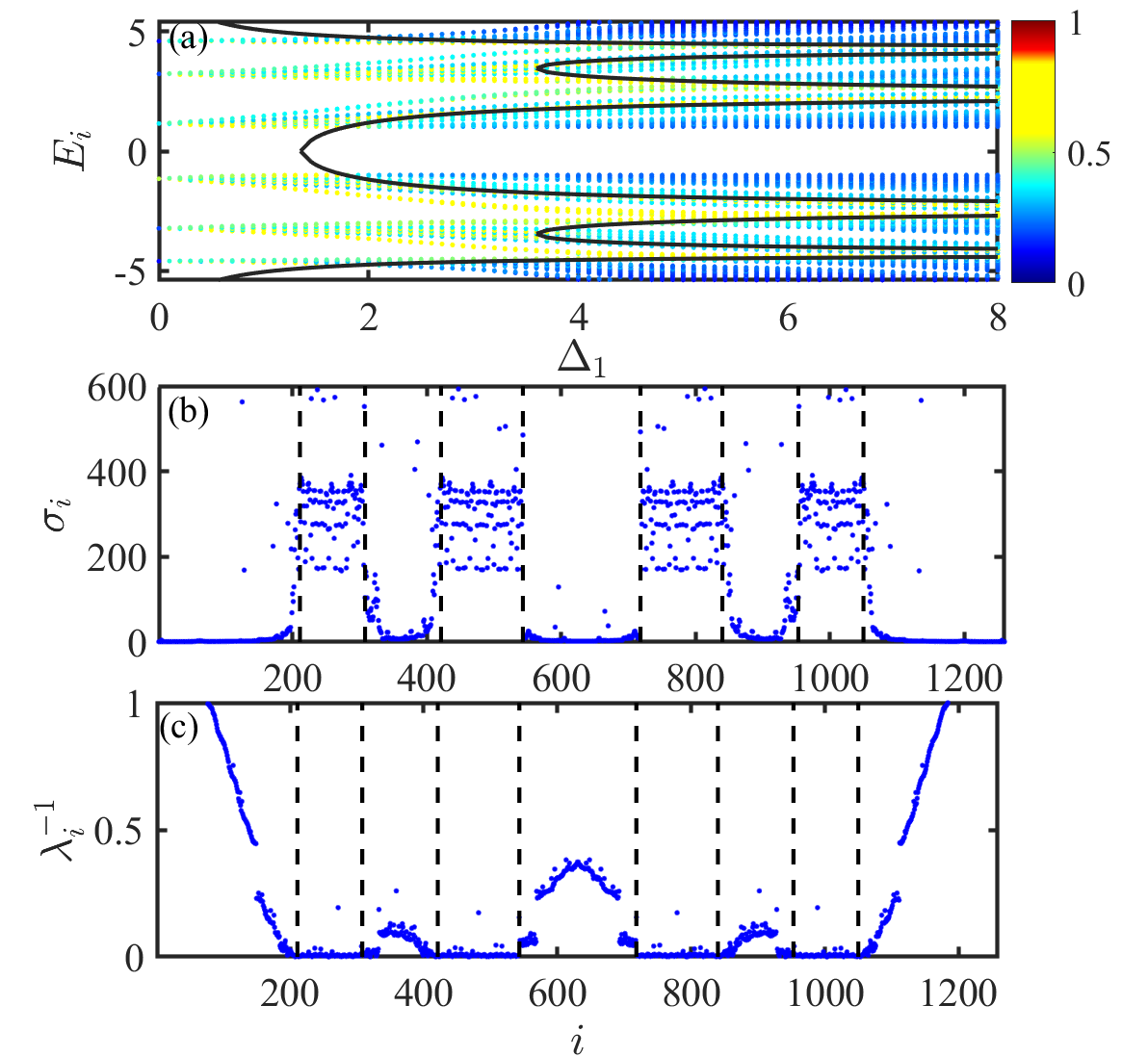}
		\caption{(a) The {fractal} dimension $\mathrm{\Gamma}_i$ of different eigenstates as a function of $E_i$ and $\Delta_1$ for $L=1260$ and $\Delta_2=3$. The {black} lines represent the {AMEs} given in Eq. (\ref{eq22}). The distributions of the eigenstate coordinates $\sigma_i$ and Lyapunov exponent $\lambda^{-1}_i$ with $\Delta_1=6$ are shown in (b) and (c), respectively. The black dash lines represent the {AMEs} given by Eq. (\ref{eq22}) in (b) and (c).}
		\label{Fig5}
	\end{figure}
	
	For an arbitrary $\kappa$, the exact expression for {AMEs} (with a detailed derivation provided in the Appendix) is given by:
	\begin{eqnarray}
		2\left|\frac{(-2)^{\kappa}(\Delta_2)^{\kappa-1}\Delta_1}{\prod_{m}^{\kappa-1}A_m}\right|&=&\left|\left(\frac{4}{A_{\kappa-1}}+1\right)\Delta_1^2\right|,
		\label{eq21}
	\end{eqnarray}
	where $A_m=A_1-\frac{(-2\Delta_2)^2}{A_{m-1}}$. For $\kappa=3$, Eq. (\ref{eq21}) simplifies to:
	\begin{equation}
		E_c=\pm\sqrt{2+t^2+\Delta_2^2\pm\sqrt{4+4\Delta_2^2\pm\frac{16}{\Delta_1}\Delta_2^2}}.\label{eq22}
	\end{equation}
	{Multiple {AMEs} emerge in Fig. \ref{Fig5}(a) marked by the {black} solid lines}, where $\Gamma_i$ is shown as a function of the eigenenergies and $\Delta_1$ for {$L=1260$}. As predicted by the analytical results, $\Gamma_i$ approximately changes from $0$ to $0.5$ when the energies across these red lines. To illustrate the intermediate phase with {AMEs}, we take $\Delta_1=6$ as a specific example. In Figs. \ref{Fig5}(b) and \ref{Fig5}(c), we show the values of $\sigma_i$ and  $\lambda_i$ for different eigenstates, respectively. The black dash lines represent the {AMEs} given by Eq. (\ref{eq22}). In the critical regions, $\sigma_i$ exhibits significant fluctuation, and the corresponding values of $\lambda^{-1}_i$ approach $0$, indicating delocalized behavior. In contrast, in the localized regions, $\lambda^{-1}_i>0$ and $\sigma_i$ is very small, reflecting the localized properties. Moreover, for any $\kappa$, one can obtain $4(\kappa-1)$ {AMEs} described by Eq. (\ref{eq21}), and the critical regions will persist even in the large disorder regime for $\kappa\ge2$.
			\begin{figure}[h]
		\includegraphics[width=0.52\textwidth]{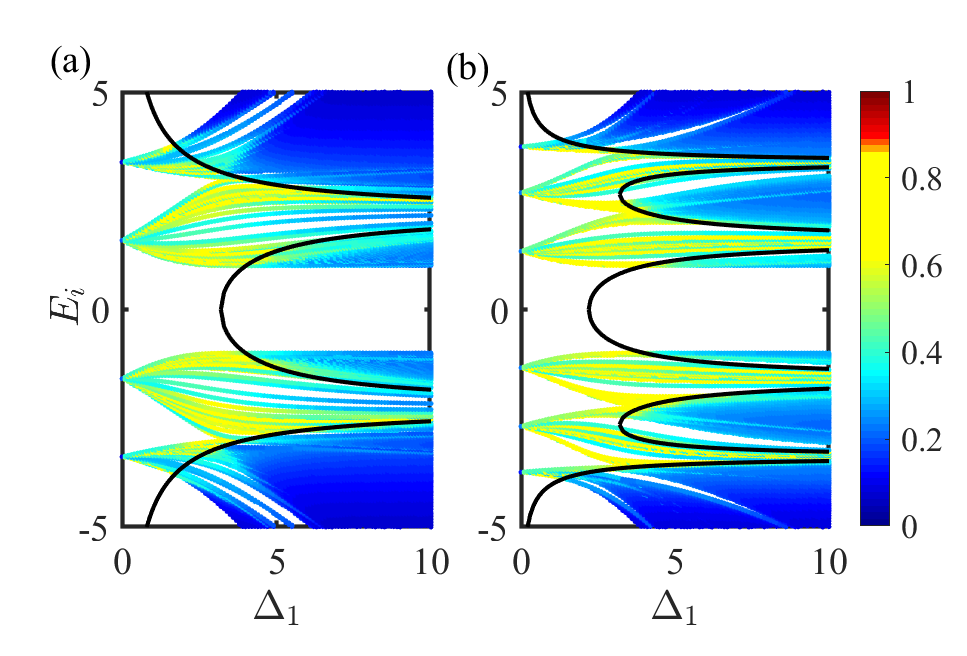}
		\caption{{The fractal dimension $\Gamma_i$ of different eigenstates as a function of the corresponding $E_i$ and $\Delta_1$ with small random perturbations in intracell hopping for (a) $\kappa=2$ and (b) $\kappa=3$. The {black} lines represent the {AMEs} given in Eq. (\ref{eq22}) and Eq. (\ref{eq19}) without small random perturbations in (a) and (b). Other parameters: {$L=1260$} and $\Delta_2=2$.} }
		\label{Fig6_1}
	\end{figure}

{To clearly demonstrate the robustness of {AMEs} even if flat lattice lattice structure is disrupted, we introduce small random perturbations in intracell hopping and plot the fractal dimension $\Gamma_i$ of different eigenstates as a function of the corresponding $E_i$ and $\Delta_1$ or (a) $\kappa=2$ and (b) $\kappa=3$ in Fig \ref{Fig6_1}(a) and (b). Specifically, we use a random disorder amplitude set to $10^{-4}$ in Fig \ref{Fig6_1}. Due to the emergence of the random perturbations, the ideal flat lattice structure will be destroyed. From Fig \ref{Fig6_1}, one can see that the {AMEs} still exist in the system. However, the analytical solution of {AMEs} can no longer accurately determine their positions, indicating that the random perturbation causes a shift in the positions of {AMEs} within the system. These  results indicate that {AMEs} are robust against small perturbations.}
	\section{Experimental realization}
	
	\begin{figure}[tbp]
		\begin{center}
			\includegraphics[width=0.5\textwidth]{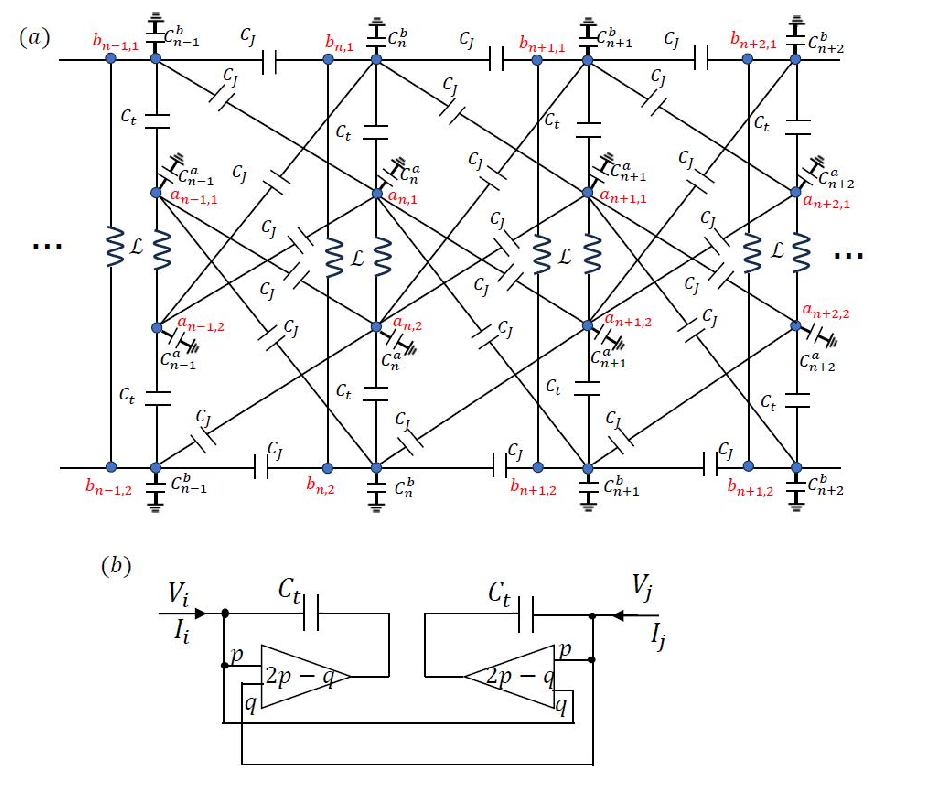}
		\end{center}
		\caption{(Color online) (a) Electrical circuit implementation of the system. {(b) The
equivalent negative impedance between two free terminals, where the markings on the ideal amplifier represent the relationship
between output voltage and input voltage.}
			\label{Fig6}}
	\end{figure}
	
	This mosaic modulated cross-stitch lattice with multiple {AMEs} can be experimentally realized in electrical circuits. We designed a electrical circuit in Fig. \ref{Fig6}(a), corresponding to the model described by $\hat{H} = \hat{H}_0+\hat{H}^{\prime}$. In Fig. \ref{Fig6}(a), the intracell and the intercell hopping amplitudes can be controlled by the capacitor $C_t$, $C_J$, respectively. {We can use a two-terminal configuration to achieve the negative capacitor value $-C_t$, which consists of two capacitors and two operational amplifiers \cite{EHCAPS1,EHCAPS2,LLH2024}, shown in Fig. \ref{Fig6}(b).} The inductor $\mathcal{L}$ is	used to tune the resonant frequency of the circuit, and the on-site potentials at each lattice site is simulated by grounding capacitors $C_n^a, C_n^b$. The whole system can be represented by the circuit Laplacian $\mathcal{J}(\omega)$ of the circuit. The Laplacian is defined as the response of the grounded-voltage vector $\mathcal{V}$ to the vector $I$ of input current by \cite{LLH2024,LCH2019PRL}
	\begin{equation}
		I(\omega)=\mathcal{J}(\omega)\mathcal{V}(\omega).\label{eq26}
	\end{equation}
	Using Eq. (\ref{eq26}), the current of each node within the unit cell can be expressed as	
	\begin{eqnarray}
		{I}_{n}^a&=&i\omega C_t \mathcal{I}_2\mathcal{V}_{n}^b+i\omega C_J\begin{pmatrix}0&1\\1&0\end{pmatrix} \Big[\mathcal{V}_{n+1}^a+\mathcal{V}_{n+1}^b+\mathcal{V}_{n-1}^a\Big]\notag\\
		&&+i\omega C_J\mathcal{I}_2\mathcal{V}_{n-1}^b+\Big[\frac{1}{i\omega \mathcal{L}}\begin{pmatrix}-1&1\\1&-1\end{pmatrix}-i\omega \Big(C_{n,a}+C_t\notag\\
		&&+4C_J\Big)\mathcal{I}_2\Big]\mathcal{V}_{n}^a \label{eq27}
	\end{eqnarray}
	and
	\begin{eqnarray}
		{I}_{n}^b&=&i\omega C_t \mathcal{I}_2\mathcal{V}_{n}^a+i\omega C_J\begin{pmatrix}0&1\\1&0\end{pmatrix}\mathcal{V}_{n-1}^a+i\omega C_J\mathcal{I}_2\Big[\mathcal{V}_{n+1}^b\notag\\
		&+&\mathcal{V}_{n+1}^a+\mathcal{V}_{n-1}^b\Big]+\Big[\frac{1}{i\omega \mathcal{L}}\begin{pmatrix}-1&1\\1&-1\end{pmatrix}-i\omega \Big(C_{n,b}+C_t\notag\\
		&+&4C_J\Big)\mathcal{I}_2\Big]\mathcal{V}_{n}^{{b}},\label{eq28}
	\end{eqnarray}
	where the vector $I_{n}^{\overline{\alpha}}=(I_{n,1}^{\overline{\alpha}},I_{n,2}^{\overline{\alpha}})^T$ and $\mathcal{V}_{n}^{\overline{\alpha}}=(\mathcal{V}_{n,1}^{\overline{\alpha}},\mathcal{V}_{n,2}^{\overline{\alpha}})^T$ with $\overline{\alpha}=\{a,b\}$ represent the node currents and voltages of $a$ and $b$ sublattices within the $n$th unit cell, respectively, and $\mathcal{I}_2$ represents the $2\times 2$ identity matrix. We transform Eqs. (\ref{eq27}) and (\ref{eq28}) using a unitary matrix
	\begin{eqnarray}
		U&=&\frac{1}{\sqrt{2}}\left(\begin{array}{cc}1&1\\1&-1\end{array}\right),\label{eq29}
	\end{eqnarray}
	and achieve the transformed current-voltage relationship
	\begin{eqnarray}
		\tilde{I}_{n}^a&=&i\omega C_t \mathcal{I}_2\tilde{\mathcal{V}}_{n}^b+i\omega C_J\begin{pmatrix}1&0\\0&-1\end{pmatrix} \Big[\tilde{\mathcal{V}}_{n+1}^a+\tilde{\mathcal{V}}_{n+1}^b\notag\\
		& &+\tilde{\mathcal{V}}_{n-1}^a\Big]+i\omega C_J\mathcal{I}_2\tilde{\mathcal{V}}_{n-1}^b+\Big[\frac{1}{i\omega \mathcal{L}}\begin{pmatrix}0&0\\0&-2\end{pmatrix}\notag\\
		& &-i\omega \Big(C_{n,a}+C_t+4C_J\Big)\mathcal{I}_2\Big]\tilde{\mathcal{V}}_{n}^a
			\label{eq30}
	\end{eqnarray}
	and
	\begin{eqnarray}
		\tilde{I}_{n}^b&=&i\omega C_t \mathcal{I}_2\tilde{\mathcal{V}}_{n}^a+i\omega C_J\begin{pmatrix}1&0\\0&-1\end{pmatrix}\tilde{\mathcal{V}}_{n-1}^a+i\omega C_J\mathcal{I}_2\notag\\
		&&\Big[\tilde{\mathcal{V}}_{n+1}^b\notag+\tilde{\mathcal{V}}_{n+1}^a+\tilde{\mathcal{V}}_{n-1}^b\Big]+\Big[\frac{1}{i\omega \mathcal{L}}\begin{pmatrix}0&0\\0&-2\end{pmatrix}\notag\\
		&&-i\omega \Big(C_{n,b}+C_t+4C_J\Big)\mathcal{I}_2\Big]\tilde{\mathcal{V}}_{n}^b.
			\label{eq31}
	\end{eqnarray}
	As shown in  Eqs. (\ref{eq30}) and (\ref{eq31}), the intercell hopping $-J$ appears in the second node of the circuit. We rewrite the current-voltage equations for the second nodes as
	\begin{eqnarray}
		\tilde{I}_{n,2}^a&=&i\omega C_t\tilde{\mathcal{V}}_{n,2}^b-i\omega C_J \Big[\tilde{\mathcal{V}}_{n+1,2}^a+\tilde{\mathcal{V}}_{n+1,2}^b\notag\\
		&&+\tilde{\mathcal{V}}_{n-1,2}^a\Big]+i\omega C_J\tilde{\mathcal{V}}_{n-1,2}^b+i\omega \Big[\frac{2}{\omega^2 \mathcal{L}}-\notag\\
		&&\Big(C_{n,a}+C_t+4C_J\Big)\Big]\tilde{\mathcal{V}}_{n,2}^a
			\label{eq32}
	\end{eqnarray}
	and
	\begin{eqnarray}
		\tilde{I}_{n,2}^b&=&i\omega C_t\tilde{\mathcal{V}}_{n,2}^a-i\omega C_J\tilde{\mathcal{V}}_{n-1,2}^a+i\omega C_J\Big[\tilde{\mathcal{V}}_{n+1,2}^b\notag\\
		&&+\tilde{\mathcal{V}}_{n+1,2}^a+\tilde{\mathcal{V}}_{n-1,2}^b\Big]+i\omega\Big[\frac{2}{\omega^2 \mathcal{L}}-\Big(C_{n,b}+C_t\notag\\
		&&+4C_J\Big)\Big]\tilde{\mathcal{V}}_{n,2}^b,
			\label{eq33}
		\end{eqnarray}
	where $\tilde{I}_{n,2}^{\overline{\alpha}}$ and $\tilde{\mathcal{V}}_{n,2}^{\overline{\alpha}}$ denote the current and voltage of the second node within each sublattice in circuit, respectively. Therefore, we obtain the targeted circuit Laplacian $\mathcal{J}(\omega)$ to simulate the our model $\hat{H} = \hat{H}_0+\hat{H}^{\prime}$ as
	\begin{small}
		\begin{align}
			\mathcal{J}(\omega)&=i\omega \begin{pmatrix}
				-C_{1,a} & C_t & -C_J & -C_J  & 0 & \dots & 0\\
				C_t & -C_{1,b} & C_J  & C_J  & 0& \dots & 0\\
				-C_J  & C_J  & -C_{2,a} & C_t & -C_J & \dots & 0\\
				-C_J  & C_J  &{C_t} & -C_{2,b} & C_J & \dots & 0\\
				\vdots & \vdots & \vdots & \ddots & \vdots & \ddots & \vdots\\
				0 & 0 & 0 & 0 & \dots & -C_{n,a} & C_t\\
				0 & 0 & 0 & 0 & \dots & C_t & -C_{n,b}\end{pmatrix}\nonumber\\
			&+\Big(\frac{2}{(\omega)^2 \mathcal{L}}-C_t-4C_J\Big){\mathcal{I}_{2n}},
			\label{eq34}
		\end{align}
	\end{small}
	where $\mathcal{I}_{2n}$ represents the $2n\times 2n$ identity matrix. {Hence the Hamiltonian for the cross-stitch flat band lattice with an on-site mosaic quasi-periodic modulation is achieved. The energy spectrum of the system can be obtained from the admittance spectrum of the circuit, and the
distribution of states can be detected by measuring the voltage at each node \cite{EHCAPS1,EHCAPS2,LCH2019PRL,LCH2018NP}.}

	\section{Summary}
	In this study, we investigated a class of exact {AMEs} in a cross-stitch lattice with quasi-periodic mosaic modulations. When $\Delta_2=0$, the system remains in a localized phase regardless of the disorder strength. However, when $\Delta_2\neq0$, we analytically demonstrate the existence of {AMEs} and derive their expressions, which show excellent agreement with numerical results. These {AMEs} are symmetrically distributed in energy spectrum and persist even in the presence of strong disorder. Moreover, we propose an experimental realization using electrical circuits, providing a practical platform for realizing our system. Our findings offer new insights into {AMEs} in flat band lattice.

	\section*{Appendix: Arbitrary $\kappa$ case for $\Delta_2\ne 0$}
	\setcounter{equation}{0}
	\setcounter{figure}{0} \setcounter{table}{0} %
	\renewcommand{\theequation}{A\arabic{equation}}
	\renewcommand{\thefigure}{A\arabic{figure}}
	\renewcommand{\thetable}{A\Roman{table}}
	
	We begin by considering the case of $\kappa=3$. In this case, the on-site potential is quasi-periodic at the $3s$-th unit cells, while at other unit cells, it remains a constant with a potential $\Delta_2\neq 0$. From Eq. (\ref{eq8}), the effective eigenvalue equations for the P-chain can be expressed as:
	\begin{align}
		A_1p_{1,s}&=-2V_{3,s-1}p_{3,s-1}-2V_{1,s}p_{2,s}\label{eqA_2} \\
		A_1p_{2,s}&=-2V_{1,s}p_{1,s}-2V_{2,s}p_{3,s}\label{eqA_3} \\
		B_0p_{3,s}&=-2V_{2,s}p_{2,s}-2V_{3,s}p_{1,s+1}+V_{3,s}^2p_{3,s}\label{eqA_4}.
	\end{align}
    Here, $A_1=B_0-\Delta_2^2$. Combining Eqs. (\ref{eqA_2}) and (\ref{eqA_3}), we obtain:
	\begin{align}
		\left[A_1-\frac{\left(-2\Delta_2\right)^2}{A_1}\right]p_{1,s}&=-2V_{3,s-1}p_{3,s-1}+\frac{\left(-2\Delta_2\right)^2}{A_1}p_{3,s},\label{eqA_5}\\
		\left[A_1-\frac{\left(-2\Delta_2\right)^2}{A_1}\right]p_{2,s}&=-2\frac{-2\Delta_2}{A_1}V_{3,s-1}p_{3,s-1}-2\Delta_2p_{3,s}\label{eqA_6}.
	\end{align}
	Next, by combining the results of (\ref{eqA_4}), (\ref{eqA_5}) and (\ref{eqA_6}), we arrive at the following equation:
	\begin{eqnarray}
		\left(B_0-\frac{4\Delta_2^2}{A_2}\right)p_{3,s}&=&-\frac{8\Delta_2^2}{A_1A_2}V_{3,s-1}p_{3,s-1}-\frac{8\Delta_2^2}{A_1A_2}V_{3,s}p_{3,s+1}\notag\\
		&+&\left(\frac{4}{A_2}+1\right)V_{3,s}^2p_{3,s}\label{eqA_7},
		\label{eqA_7}
	\end{eqnarray}
	where $A_2=A_1-\frac{\left(-2\Delta_2\right)^2}{A_1}$. The eigenvalue equation (\ref{eqA_7}) corresponds to a generalized AA model with diagonal and off-diagonal quasi-periodic modulations, where exhibits a critical-to-insulator transition. Based on the localization transition point of the generalized AA model \cite{WYCLXJ2021,Lee2023}, the expression for the {AMEs} is given by:
	\begin{eqnarray}
		2\left|\frac{-8\Delta_2^2\Delta_1}{A_1A_2}\right|&=&\left|\left(\frac{4}{A_2}+1\right)\Delta_1^2\right|.
		\label{eqA_8}
	\end{eqnarray}
	Simplifying Eq.(\ref{eqA_8}), we obtain {AMEs}
	\begin{eqnarray}
		E_c&=&\pm\sqrt{2+t^2+\Delta_2^2\pm\sqrt{4+4\Delta_2^2\pm\frac{16}{\Delta_1}\Delta_2^2}},
		\label{eqA_9}
	\end{eqnarray}
    which are represented by the {black} solid lines in Fig. \ref{Fig5}(a).
	\begin{figure}[h]
		\includegraphics[width=0.5\textwidth]{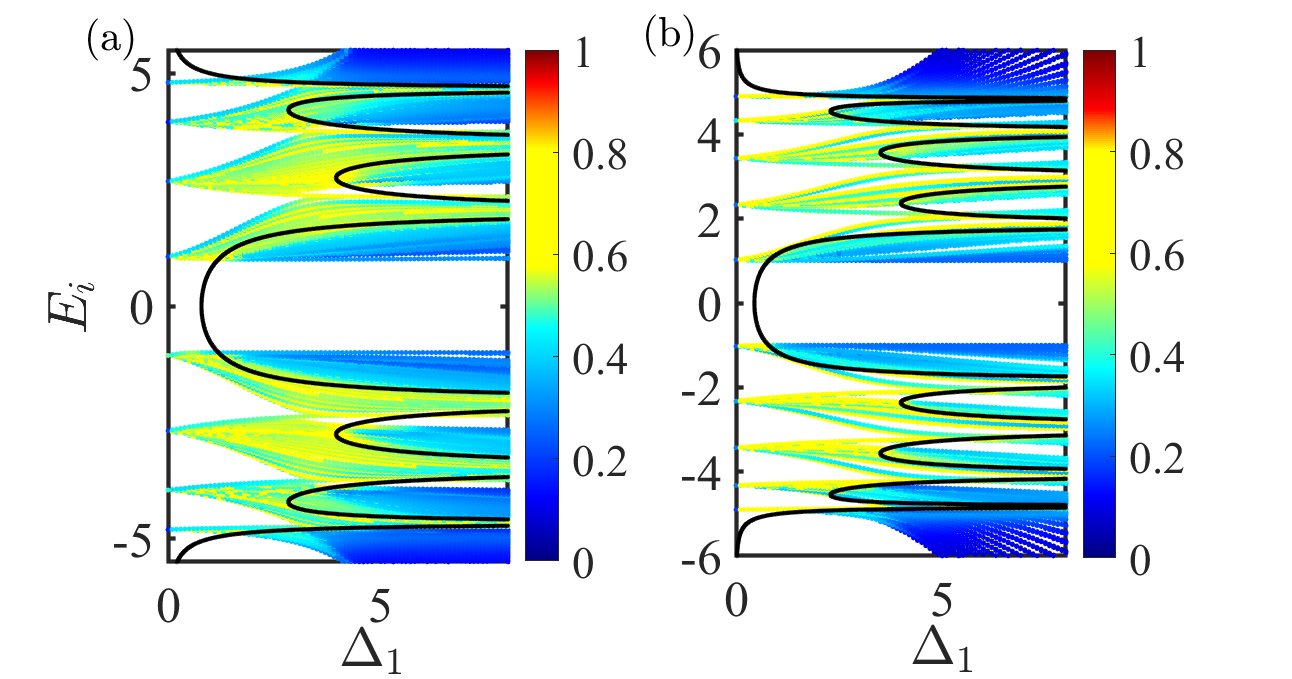}
		\caption{The Fractal dimension $\mathrm{\Gamma}_i$ of different eigenstates as a function of corresponding $E_i$ and $\Delta_1$ for (a) $\kappa=4$ and (b) $\kappa=5$, respectively. The black lines represent the exact {AMEs} given in Eq.(\ref{eqA_12}). Here, $N=800$ and $\Delta_2=3$.}
		\label{Fig7}
	\end{figure}
	For any $\kappa$, the effective eigenvalue equations for the P-chain can be expressed as:
	\begin{small}
		\begin{eqnarray}
			A_1p_{1,s}&=&-2V_{\kappa,s-1}p_{\kappa,s-1}-2V_{1,s}p_{2,s}\notag \\
			&...&\notag \\
			B_0p_{\kappa,s}&=&-2V_{\kappa-1,s}p_{\kappa-1,s}-2V_{\kappa,s}p_{1,s+1}+V_{\kappa,s}^2p_{\kappa,s}.
			\label{eqA_10}
		\end{eqnarray}
	\end{small}
	Following a similar process as in the $\kappa=2$ and $3$ cases, and employing inductive reasoning, we derive:
	\begin{align}
		B_0p_{\kappa,s}&=\frac{(-2)^{\kappa}(\Delta_2)^{\kappa-1}}{\prod_{m}^{\kappa-1}A_m}\left(V_{\kappa,s-1}p_{\kappa,s-1}+V_{\kappa,s}p_{\kappa,s+1}\right)\\
		&+\frac{(-2\Delta_2)^2}{A_{\kappa-1}}p_{\kappa,s}\notag+\left(\frac{4}{A_{\kappa-1}}+1\right)V_{\kappa,s}^2p_{\kappa,s}.
		\label{eqA_11}
	\end{align}
	The corresponding expression for the {AMEs} is:
	\begin{eqnarray}
		2\left|\frac{(-2)^{\kappa}(\Delta_2)^{\kappa-1}\Delta_1}{\prod_{m}^{\kappa-1}A_m}\right|&=&\left|\left(\frac{4}{A_{\kappa-1}}+1\right)\Delta_1^2\right|,
		\label{eqA_12}
	\end{eqnarray}
    where $A_m=A_1-\frac{(-2\Delta_2)^2}{A_{m-1}}$.
	To verify the accuracy of these results, we plot $\mathrm{\Gamma}_i$ as a function of the eigenenergies $E_i$ and $\Delta_1$ for $\kappa=4$ and $5$, as shown in Fig. \ref{Fig7}(a) and \ref{Fig7}(b). The black lines represent the analytic results from Eq. (\ref{eqA_12}). The analytical and numerical results are in good agreement.
	\section*{Acknowledgments}
	Zhihao Xu is supported by the NSFC (Grant No. 12375016) and Beijing National Laboratory for Condensed Matter Physics (No. 2023BNLCMPKF001). Yunbo Zhang is supported by the NSFC (Grant No. 12474492 and 12461160324). This work is also supported by NSF for Shanxi Province (Grant No. 1331KSC).


\begin{references}		
		%%critical states
        \bibitem{Huckestein}{B. Huckestein, Rev. Mod. Phys. \textbf{67,} 357 (1995).}
		\bibitem{Jagannathan}{A. Jagannathan, Rev. Mod. Phys. \textbf{93,} 045001 (2021).}		
		\bibitem{Ostlund}{S. Ostlund, R. Pandit, D. Rand, H. J. Schellnhuber, and E. D. Siggia,  Phys. Rev. Lett. \textbf{50}, 1873 (1983).}
		\bibitem{Merlin}{R. Merlin, K. Bajema, R. Clarke, F. Y. Juang, and P. K. Bhattacharya,  Phys. Rev. Lett. \textbf{55}, 1768 (1985).}
		\bibitem{E}{E. Maci\'{a}, Phys. Rev. B \textbf{60}, 10032 (1999).}
		\bibitem{Ashraff}{J. A. Ashraff and R. B. Stinchcomebe, Phys. Rev. B \textbf{40}, 2278 (1989).}
        \bibitem{Evers}{F. Evers and A. D. Mirlin, Rev. Mod. Phys. \textbf{80,} 1355 (2008).}
        \bibitem{Xiao2021SC}{T. Xiao, D. Xie, Z. Dong, T. Chen, W. Yi, and B. Yan, Sci. Bull. \textbf{66,} 2175 (2021).}
		\bibitem{XXCPL2024}{T. Liu and X. Xia, Chin. Phys. Lett. \textbf{41,} 017102 (2024).}
        \bibitem{Goblot2020}{V. Goblot, A. \v{S}trkalj, N. Pernet, J. L. Lado, C. Dorow, A. Lema\^{\i}tre, L. Le Gratiet, A. Harouri, I. Sagnes, S. Ravets, A. Amo, J. Bloch and O. Zilberberg, Nat. Phys. {\bf16}, 832-836 (2020).}
        \bibitem{ZZhai2021}{L. J. Zhai, G. Y. Huang, S. Yin, Phys. Rev. B. \textbf{104,} 014202 (2021).}
         \bibitem{Daiprb2024}{Q. Dai, Z. Lu, and Z. Xu, Phys. Rev. B \textbf{108,} 144207 (2023).}
        \bibitem{LTarxiv2024}{T. Liu, arXiv:2411.09067.}
        \bibitem{Lagendijk}{A. Lagendijk, B. van Tiggelen, and D. S. Wiersma, Phys. Today \textbf{62,} 24 (2009).}
        %%fractial dimension
        \bibitem{LinPRB2023}{X. Lin, X. Chen, G.-C. Guo, and M. Gong, Phys. Rev. B \textbf{108,} 174206 (2023).}
        \bibitem{SRoy2021PRL}{S. Roy, T. Mishra, B. Tanatar, and S. Basu, Phys. Rev. Lett. \textbf{126,} 106803 (2021).}
        \bibitem{Padhan2022PRB}{A. Padhan, M. K. Giri, S. Mondal and T. Mishra,  Phys. Rev. B \textbf{105,} L220201 (2022).}
        \bibitem{ZWZPRA2022}{Z.-W. Zuo and D. Kang, Phys. Rev. A \textbf{106,} 013305 (2022).}
        \bibitem{SA2023PRB}{S. Aditya, K. Sengupta, and D. Sen, Phys. Rev. B \textbf{107,} 035402 (2023).}
        \bibitem{SRoy2022PRB}{S. Roy, S. Chattopadhyay, T. Mishra, and S. Basu, Phys. Rev. B \textbf{105,} 214203 (2022).}
		%critical phase
		\bibitem{TLT2021}{L.-Z. Tang, G.-Q. Zhang, L.-F. Zhang, and D.-W. Zhang, Phys. Rev. A \textbf{103,} 033325 (2021).}
		\bibitem{Duncan2024}{C. W. Duncan, Phys. Rev. B \textbf{109,} 014210 (2024).}
		\bibitem{WYC2024}{C. Yang, W. Yang, Y. Wang and Y. Wang, Phys. Rev. A \textbf{110,} 042205 (2024).}
		\bibitem{WYC2020}{Y. Wang, L. Zhang, S. Niu, D. Yu, X.-J. Liu, Phys. Rev. Lett. \textbf{125,} 073204 (2020).}
		\bibitem{Yicaizhang2022}{Y.-C. Zhang and Y.-Y. Zhang, Phys. Rev. B \textbf{105,} 174206 (2022).}
		\bibitem{WYCLXJ2021}{Y. Wang, C. Cheng, X.-J. Liu, and D. Yu, Phys. Rev. Lett. \textbf{126,} 080602 (2021).}
		\bibitem{Ribeiro2}{M. Gon\c{c}alves, B. Amorim, F. Riche, E. V. Castro, and P. Ribeiro,  arXiv:2305.03800.}
		%\TEXTCOLOR{RED}{AMES}
		\bibitem{TL2023}{T. Liu, X. Xia, S. Longhi, and L. Sanchez-Palencia, Sci-Post Phys.  \textbf{12,} 27 (2022).}
		\bibitem{XJL2023}{X.-C. Zhou, Y. Wang, T.-F. J. Poon, Q. Zhou, and X.-J. Liu, Phys. Rev. Lett. \textbf{131,} 176401 (2023).}
        {\bibitem{LSZ2025}{S.-Z. Li, Y.-C. Zhang, Y.-C. Wang, S.C. Zhang, S.-L. Zhu, and Z. Li, arXiv:2501.07866(2025).}}
        {\bibitem{CYD2015}{F. Liu, S. Ghosh, and Y. D. Chong, Phys. Rev. B \textbf{91,} 014108 (2015).}}
        %MEs
		\bibitem{Kraus2012PRL}{Y. E. Kraus, Y. Lahini, Z. Ringel, M. Verbin, and O. Zilberberg, Phys. Rev. Lett. \textbf{109,} 106402 (2012).}
		\bibitem{Segev2013NP}{M. Segev, Y. Silberberg, and D. N. Christodoulides, Nat. Photon. \textbf{7,} 197 (2013).}
		\bibitem{Biddle2010PRL}{J. Biddle and S. Das Sarma, Phys. Rev. Lett. \textbf{104,} 070601 (2010).}
		\bibitem{Ganeshan2015PRL}{S. Ganeshan, J. H. Pixley, and S. Das Sarma, Phys. Rev. Lett. \textbf{114,} 146601 (2015).}
		\bibitem{XJL2022}{Y. Wang, L. Zhang, W. Sun, T.-F. J. Poon, and X.-J. Liu,  Phys. Rev. B \textbf{106,} L140203 (2022).}
		\bibitem{YW2023}{Y. Wang, L. Zhang, Y. Wan, Y. He, and Y. Wang, Phys. Rev. B \textbf{107,} L140201 (2023).}
		\bibitem{LZP2022}{Z. Lu, Z. Xu, and Y. Zhang, Ann. Phys. (Berlin) \textbf{534,} 2200203 (2022).}
		\bibitem{ZHXU2020}{Z. Xu, H. Huangfu, Y. Zhang, and S. Chen, New J. Phys. \textbf{22,} 013036 (2020).}
		\bibitem{LLH2024}{H. Liu, Z. Lu, Xu X and Z. Xu, New J. Phys. \textbf{26,} 093007 (2024).}
		\bibitem{Zhihaoxu2021}{Z. Xu, X. Xia, and S. Chen, Phys. Rev. B \textbf{104,} 224204 (2021).}
		\bibitem{Longhi2024}{S. Longhi, Phys. Rev. Lett. \textbf{132,} 236301 (2024).}
		\bibitem{LT2020}{T. Liu, H. Guo, Y. Pu, and S. Longhi, Phys. Rev. B \textbf{102,} 024205 (2020).}
		\bibitem{LT2018}{T. Liu and H. Guo, Phys. Rev. B \textbf{98,} 104201 (2018).}
		\bibitem{Longhi2024PRB}{S. Longhi, Phys. Rev. B \textbf{110,} 184201 (2024).}
        \bibitem{LuschenPRL}{H. P. Luschen, S. Scherg, T. Kohlert, M. Schreiber, P. Bordia, X. Li, S. Das Sarma, and I. Bloch, Phys. Rev. Lett. \textbf{120,} 160404 (2018).}
		\bibitem{SarmaPRB2020}{X. Li and S. Das Sarma, Phys. Rev. B \textbf{101,} 064203 (2020).}
        \bibitem{Bodyfelt2014}{J. D. Bodyfelt, D. Leykam, C. Danieli, X. Yu, and S. Flach, Phys. Rev. Lett. \textbf{113,} 236403 (2014).}
        \bibitem{YaoPRL2020}{H. Yao, T. Giamarchi, and L. Sanchez-Palencia, Phys. Rev. Lett. \textbf{125,} 060401 (2020).}
        \bibitem{HY2020PRL}{Y. Wang, X. Xia, L. Zhang, H. Yao, S. Chen, J. You, Q. Zhou, and X.-J. Liu, Phys. Rev. Lett. \textbf{125,} 196604 (2020).}
        \bibitem{GaoJun2}{J. Gao, I. M. Khaymovich, A. Iovan, X.-W. Wang, G. Krishna, Z.-S. Xu, E. Tortumlu, A. V. Balatsky, V. Zwiller, and A. W. Elshaari, Phys. Rev. B \textbf{108,} L140202 (2023).}
        %%%add mobility edges
        {\bibitem{LZP2025FOP}{Z. Lu, Y. Zhang, and Z. Xu, Front. Phys. \textbf{20,} 024204 (2025).}}
        {\bibitem{LIUTONGCPB1}{T. Liu, S. Cheng, R. Zhang, R. Ruan, and H. Jiang, Chin. Phys. B \textbf{31,} 027101 (2022).}}
        {\bibitem{LIUTONGCPB2}{T. Liu and S. Cheng, Chin. Phys. B \textbf{32,} 027102 (2023).}}
        {\bibitem{EHCAPS1}{C. H. Xu, Z. C. Xu, Z. Y. Zhou, E. H. Cheng, and L. J. Lang, Acta Phys. Sin. \textbf{72,} 200301 (2023).}}
        {\bibitem{EHCAPS2}{E. H. Cheng and L. J. Lang, Acta Phys. Sin. \textbf{71,} 160301 (2022).}}
        %%%add papers
        {{\bibitem{paper1}{D. S. Borgnia, A. Vishwanath, and R.-J. Slager, Phys. Rev. B \textbf{106,} 054204 (2022).}}}
        {{\bibitem{paper2}{D. S. Borgnia and R.-J. Slager,Phys. Rev. B \textbf{107,} 085111 (2023).}}}
        %%%%flat experiment
        \bibitem{Bilal2024PRL}{M. M. Samak and O. R. Bilal, Phys. Rev. Lett. \textbf{133,} 266101 (2024).}
        %%%%flat band
        \bibitem{Maksymenko2012}{M. Maksymenko, A. Honecker, R. Moessner, J. Richter, and O. Derzhko,  Phys. Rev. Lett. \textbf{109,} 096404 (2012).}
		\bibitem{Rhim2021}{J.-W. Rhim and B.-J. Yang, Adv. Phys.: X \textbf{6,} 1901606 (2021).}
		%%%%引起平带的方法
		\bibitem{Hyr2013}{M. Hyrk\"{a}s, V. Apaja, and M. Manninen, Phys. Rev A \textbf{87,} 023614 (2013).}
		\bibitem{Derzhko2010}{O. Derzhko, J. Richter, A. Honecker, M. Maksymenko, and R. Moessner, Phys. Rev. B \textbf{81,} 014421 (2010).}
		\bibitem{Goda2006}{M. Goda, S. Nishino, and H. Matsuda, Phys. Rev. Lett. \textbf{96,} 126401 (2006).}
		\bibitem{Huber2010}{S. D. Huber and E. Altman, Phys. Rev. B \textbf{82,} 184502 (2010).}
		\bibitem{Green2010}{D. Green, L. Santos, and C. Chamon, Phys. Rev. B \textbf{82,} 075104 (2010).}
        %%%added flat band
        {\bibitem{FOPFLAT1}{X. Li, J. Liu, and T. Liu, Front. Phys. \textbf{19,} 33211 (2024).}}
        {\bibitem{FOPFLAT2}{Z. Ma, W. J. Chen, Y. Chen, J. H. Gao, and X. C. Xie, Front. Phys. \textbf{18,} 63302 (2023).}}
        {\bibitem{FOPFLAT3}{Z. Ma, S. Li, M. M. Xiao, Y. W. Zheng, M. Lu, H. Liu, J. H. Gao, and X. C. Xie, Front. Phys. \textbf{18,} 13307 (2023).}}
        \bibitem{Biao2024PRB}{Y. Biao, Z. Yan, and R. Yu, Phys. Rev. B \textbf{110,} L241110 (2024).}
        %%flat band
        \bibitem{Danieli2015}{C. Danieli, J. D. Bodyfelt, and S. Flach, Phys. Rev. B \textbf{91,} 235134 (2015).}
		\bibitem{Lee2023}{S. Lee, A. Andreanov, and S. Flach, Phys. Rev. B \textbf{107,} 014204 (2023).}
		\bibitem{Maimaiti2017}{W. Maimaiti, A. Andreanov, H. C. Park, O. Gendelman, and S. Flach, Phys. Rev. B \textbf{95,} 115135 (2017).}
		\bibitem{Maimaiti2021}{W. Maimaiti and A. Andreanov, Phys. Rev. B \textbf{104,} 035115 (2021).}
        %%Lyapunov exponent
		\bibitem{Avila}{A. Avila, Acta. Math. \textbf{215}, 1 (2015).}
		\bibitem{You}{A. Avila, J. You, and Q. Zhou, Duke. Math. J. \textbf{166}, 2697 (2017).}
        \bibitem{Shu}{Y. Liu, Q. Zhou, and S. Chen, Phys. Rev. B \textbf{104}, 024201 (2021).}
        \bibitem{Zhou}{Y. Liu, Y. Wang,  X.-J. Liu, Q. Zhou, and S. Chen, Phys. Rev. B \textbf{103}, 014203 (2021).}
		%%%transfer matrix method
		\bibitem{WZH2022PRB}{Z.-H. Wang, F. Xu, L. Li, D.-H. Xu, and B. Wang, Phys. Rev. B \textbf{105,} 024514 (2022).}
		\bibitem{MacKinnon1983}{A. MacKinnon and B. Kramer, Z. Phys. B \textbf{53,} 1 (1983).}
		\bibitem{yanxialiu2}{Y. Liu, Y. Wang, Z. Zheng, and S. Chen, Phys. Rev. B \textbf{103,} 134208 (2021).}
        \bibitem{LCH2019PRL}{T. Hofmann, T. Helbig, C. H. Lee, M. Greiter, and R. Thomale, Phys. Rev. Lett. \textbf{122,} 247702 (2019).}
        \bibitem{LCH2018NP}{S. Imhof, C. Berger, F. Bayer, J. Brehm, L. W. Molenkamp, T. Kiessling, F. Schindler, C. H. Lee, M. Greiter, T. Neupert, and R. Thomale, Nat. Phys. \textbf{14,} 925 (2018).}
	\end{references}
\end{document}